\documentclass[aps,prl,twocolumn,groupedaddress,showpacs]{revtex4}
\usepackage{graphicx}
\usepackage{amsmath}
\usepackage{amsmath}
\usepackage{amsfonts}
\usepackage{amssymb}
\usepackage{graphicx}
\usepackage{color}

\newcommand{\be}{\begin{equation}}
\newcommand{\en}{\end{equation}}

\begin{document}

\title{Dynamical scaling for critical states: is Chalker's ansatz valid for strong fractality?}


\author{V.E. Kravtsov$^1$, \ A. Ossipov$^2$, \ O.M. Yevtushenko$^3$}
\affiliation{$^1$Abdus Salam ICTP, P.O. 586, 34100 Trieste, Italy \\
                 $^2$School of Mathematical Sciences, University of Nottingham, Nottingham
                          NG7 2RD, United Kingdom \\ 
                 $^3$Arnold Sommerfeld Center and Center for Nano-Science,
                          Ludwig-Maximilians-University, Munich D-80333, Germany
                 }
\author{E. Cuevas}
\affiliation{Departamento de F\'{\i}sica, Universidad de Murcia, E-30071 Murcia, Spain}

\date{\today}

\begin{abstract}
The dynamical scaling for statistics of critical multifractal
eigenstates proposed by Chalker is analytically verified for the
critical random matrix ensemble in the limit of strong
multifractality controlled by the small parameter $b\ll 1$. The
power law behavior of the quantum return probability $P_{N}(\tau)$
as a function of the matrix size $N$ or time $\tau$ is confirmed in
the limits $\tau/N\rightarrow\infty$ and $N/\tau\rightarrow\infty$,
respectively, and it is shown that the exponents characterizing
these power laws are equal to each other up to the order $b^{2}$.
The corresponding  analytical expression for the fractal dimension
$d_{2}$ is found.
\end{abstract}



\maketitle

Fifty years after the seminal paper by P.W.Anderson \cite{PWA58}
which predicted localization it is becoming a common wisdom that the
{\it multi-fractal} scaling properties of critical \cite{Mirlin-Rew}
and near critical \cite{CK07} states are not just an exotic theory
exercise but an important physics reality \cite{Huse,Lagendijk}.
Since the pioneer's work by F.Wegner \cite{W80}, it is known that 
at the Anderson transition point  the local moments of the
wave-functions $\psi_n({\bf r})$ scale with the system size $L$ as
(summation over $n$ runs over the critical window near the mobility
edge) \be \label{IPR} I_q=\sum_{\bf r}\sum_{n}\left\langle
|\psi_n({\bf r})|^{2q}\right\rangle\propto L^{-d_q(q-1)}, \en where
$d_{q}$ is the fractal dimension corresponding to the $q$-th moment
({\it multi}-fractality). This implies \cite{Mirlin-Rew} that all
scales of the amplitude $|\psi_n({\bf r})|^{2}\sim L^{-\alpha}$ are
present with the corresponding number of sites $N_{\alpha}\sim
L^{f(\alpha)}$. The function $f(\alpha)$ known as {\it spectrum of
fractal dimensions} is given by the Legendre transform of
$\tau(q)=d_{q}(q-1)$. One can distinguish between the weak
multi-fractality where $d_{q}$ is close to the system dimensionality
$d$ and is almost linear in $q$ for not too large moments $q>1$, and
the strong multi-fractality where $d_{q}\ll d$ for all $q>1$.

Recently the interest in Anderson localization has been shifted
towards its effect for a many-body system of interacted particles.
The simplest problems of this type are the "multi-fractal"
superconductivity \cite{FIKC} and the Kondo effect \cite{Kettemann}.
For such problems the relevant quantity is the matrix element of
(local) interaction which involves wave-functions of two different
energies $E_m $ and $ E_n $: 
\be\label{mat-elem} 
K(\omega,{\bf R})=\!\! \sum_{{\bf r}} \sum_{n,m}
   \langle |\psi_{n}({\bf
      r+R})|^{2}\,|\psi_{m}({\bf r})|^{2}\,\delta(E_{m}-E_{n}-\omega)
   \rangle. 
\en 
A promising playground to observe such effects at
controllable strength of disorder and interaction are systems of
cold atoms where one-dimensional Anderson localization has been
already observed \cite{Aspect} and the observation of their two- and
three- dimensional counterparts is on the way.

It was conjectured long ago by Chalker and Daniel \cite{ChD88,
Chalk} and confirmed by numerous computer simulations
\cite{CK07,ChD88,HSch94} that at $E_{0}\gg \omega\gg \Delta$
($\Delta$ is the mean level separation): 
\be \label{Chalk-anz}
  K(\omega,0)\equiv C(\omega)\propto
  (E_{0}/\omega)^{\mu}, \ \mu=1-d_{2}/d . 
\en 
The scaling relationship Eq.(\ref{Chalk-anz}) can be viewed as a result
of application of the dynamical scaling \be
\label{dynscal}L\rightarrow L_{\omega}\propto \omega^{-\frac{1}{d}},
\en to the earlier Wegner's result \cite{Wegner-density} \be
\label{WE} K(0,R)\propto (L/R)^{d-d_{2}},\en with the simultaneous
assumption that the $R$-de\-pen\-dence is saturated for $R<\ell$; $\ell$
being the minimum scale of multi-fractality of the order of the
elastic scattering length and $E_{0}\propto \ell^{-d}$ is the
corresponding high-energy cutoff. This property is of great
importance for electron interaction in the vicinity of the Anderson
transition, as it leads to a dramatic enhancement of matrix elements
compared to the case of absence of multi-fractality
\cite{CK07,FIKC}.

Note that the dynamical scaling  hypothesis Eq.(\ref{dynscal})
implies that the energy scale $\omega$ corresponds to the length
scale $L_{\omega}$ equal to the size of a sample where the mean level spacing
is $\omega$. However simple and natural is this hypothesis, it leads
to a somewhat counter-intuitive consequence in the case of strong
multi-fractality. Indeed, in the limit $d_{2}\rightarrow 0$ the
exponent $\mu$ in Eq.(\ref{Chalk-anz}) saturates at $\mu= 1$.    It
signals of a strong overlap of two infinitely sparse fractal
wavefunctions, while one would expect such states not to overlap in
a typical realization, similar to two localized wavefunctions. At
present there is no analytical evidence of validity of
Eq.(\ref{Chalk-anz}) in the limit of strong multi-fractality though
numerical simulations seem to be very encouraging to it \cite{CK07}.
The main goal of this paper is to provide such an evidence.

To achieve this objective we consider a model system described by a
random matrix Hamiltonian with the multi-fractal eigenstates
\cite{MFD96,KM97}.  The Gaussian ensemble of such random $N\times N$
matrices is defined by the variance of random (complex) entries
\be
\label{var} v_{n-m}\equiv\left\langle |H_{nm}|^2 \right\rangle=
 \frac{\frac{1}{2}b^2}{b^{2}+|n-m|^{2(1-\epsilon)}}, \
 \langle H_{nm}\rangle=0,
\en and describes a long-range hoppings between sites of the
one-dimensional lattice. The parameter $b$ determines the strength
of the fractality, with the weak multi-fractality at $b\gg 1$ and
the strong multi-fractality at $b\ll 1$. It is the last limit that
we will concentrate on in this paper. For convenience of the further
calculations we introduced here a regularizing parameter
$\epsilon\rightarrow +0$.

Numerous computer simulations \cite{Mirlin-Rew,CK07,Kbook} show
that the above model gives an unexpectedly  good {\it quantitative}
description of statistics of the critical wave functions in the
3D Anderson model. Inspired by this fact we use the critical RMT,
Eq.(\ref{var}), as the simplest and representative
description of multi-fractality at the Anderson transition.

To proceed further on, we note that the moment $I_{2}$ defined in
Eq.(\ref{IPR}) and the dynamic correlation function $C(\omega)$ given
by Eq.(\ref{Chalk-anz}) are related with the physical observable known
as  the {\it return probability}: \be \label{RP}
P_{N}(t)=\int_{-\infty}^{\infty} d\omega\: e^{-i\omega t} C(\omega).
\en Indeed, plugging $K(\omega,0)$ defined by Eq.(\ref{mat-elem})
into Eq.(\ref{RP}) and taking into account that the average value of
$e^{-i(E_{n}-E_{m})t}\rightarrow \delta_{mn}$ in the limit
$t\rightarrow\infty$ at a fixed mean level separation $\Delta\sim
N^{-1}$ we obtain at $N\gg 1$:
\be\label{P(L)}
    \lim_{t/N\to \infty} P_{N}(t)=I_2\propto N^{-d_{2}/d} ,
\en
where in a particular case of the critical RMT the fractal dimension
$d_{2}=\sum_{n=1}^{\infty}c_{m}\,b^{m}$ allows a regular expansion
in the parameter $b\ll 1$. The first term of this expansion $c_{1}$
was computed in Ref.\cite{ME00}.

Alternatively, if $\Delta\sim N^{-1}$ tends to zero at a fixed large
$t$, the power law in Eq.(\ref{Chalk-anz}) is unbounded from below
and one obtains at $t\gg E_{0}^{-1}\sim b^{-1}$: \be\label{P(bt)}
\lim_{N/t\to \infty}P_{N}(t)\propto t^{\mu-1}\propto
t^{-d_{2}/d}.\en Thus the Chalker's ansatz Eq.(\ref{Chalk-anz}) is
equivalent to the identity:
\begin{eqnarray} \label{d-log} \left.\frac{\partial\ln
P_{N}(t)}{\partial\ln
t}\right|_{\begin{matrix}N/t\rightarrow\infty\cr
t\rightarrow\infty\end{matrix}}=
 \left.\frac{\partial\ln P_{N}(t)}{\partial\ln
N}\right|_{\begin{matrix}t/N\rightarrow\infty\cr
N\rightarrow\infty\end{matrix}}.
 \end{eqnarray}
  It implies that at large $t$ and $N$ the leading
dependence of $\ln P_{N}(t)$ on $t$ and $N$ is logarithmic, and
the coefficients in front of $\ln t$ and $\ln N$ are the same in
both limits and equal to $-d_{2}/d$. This is exactly what we are
going to demonstrate in the present work.


In order to reach this goal one needs to find the return probability
at finite time $t$ as well as its limiting value at $t\to\infty$. In
this Letter we calculate $P_{N}(t)$ using the virial expansion in
the number of resonant states, each of them is localized at a
certain site $n$. The virial expansion formalism was developed in
Ref.\cite{YK03,YO07} following the initial idea of
Ref.\cite{Levitov90}.
%
%
The supersymmetric version of the virial expansion  \cite{YO07} is
formulated in terms of  integrals over super-matrices. In
particular, it allows us to represent $P_{N}(t)$ as an infinite
series of integrals over an increasing number of
super-matrices $Q_{n}$ associated  with different sites $n$. As it was
shown in Ref.\cite{YO07}, the terms $P_{N}^{(i)}(t)$ involving
integration over $i$ different super-matrices result in the
contribution to $P_{N}(t)=\sum_{i}P_{N}^{(i)}(t)$ of the form: \be
\label{parz} P_{N}^{(i)}(t)=b^{i-1}\,p_{N}^{(i)}(bt), \en with the
known \cite{YO07} explicit expressions for $p_{N}^{(2,3)}(bt)$ and
$p_{N}^{(1)}=1$ .

It follows from Eq.(\ref{parz}) that  \be \label{parz-log}
   \ln P_{N}(t) \simeq
   b \, p_{N}^{(2)}\!(bt)+b^{2} \left[p_{N}^{(3)}\!(bt) -
   \frac{1}{2}\left(p_{N}^{(2)}\!(bt)\right)^{2} \right] .
\en
For Eq.(\ref{d-log}) to be valid, one requires that in
the limit of large $bt$, $N$: \\(i)
$p_{N}^{(2)}(bt)=-c_{1}\,\ln({\rm min}\left\{bt, N \right\})$
\\(ii)
$\left[p_{N}^{(3)}(bt)-\frac{1}{2}\left(p_{N}^{(2)}(bt)\right)^{2}
\right]=-c_{2}\,\ln({\rm min}\left\{bt, N \right\})$ \\(iii) terms
proportional to $\ln^{2}(bt)$ and $\ln^{2}N$ cancel out in the
combination (ii).

This cancelation, as well as the logarithmic asymptotic behavior
with equal coefficients in front of $\ln(bt)$ and $\ln N$ in the two
different limits, is not trivial. We show below that these
properties are hidden in the general structure of the virial
expansion and in the power law dependence of the variance of the 
critical RMT $ v_{n \ne 0} \propto 1/n^2 $, Eq(\ref{var}).

We start by analyzing the explicit expression for $P_{N}^{(2)}(t)$
obtained in Ref.\cite{YO07}: 
\be \label{p2} 
  P_{N}^{(2)}(t)=-2\sqrt{\pi}\sum_{n=1}^{N}\left\{v_{n}|t|\,e^{-v_{n}t^{2}}+\frac{\sqrt{\pi
        v_{n}}}{2}\,{\rm erf}\left( \sqrt{v_{n}}|t|\right)\right\},
\en 
where ${\rm erf}(z)=\frac{2}{\sqrt{\pi}}\int_{0}^{z}e^{-\xi^{2}}d\xi$.

In the large $t$ and $N$ limit the sum over $n$ in Eq.(\ref{p2}) is
dominated by large $n$ at any $\epsilon>0$. Replacing the sum by an
integral $\sum_{n}\rightarrow \int_{0}^{\infty}dn$
one can represent $p_{N}^{(2)}(bt)$ as a double integral: 
\be \label{double-int} 
p_{N}^{(2)}(bt)=-2 \int_{0}^{\tau}\frac{dy}{y^{2}}\int_{N^{-1}}^{\infty}\frac{dn}{n^{2}}\,{\cal
F}_{2}\left( y\, n^{1-\epsilon}\right),
\en 
where $\tau=\frac{bt}{\sqrt{2}}$, ${\cal
F}_{2}(z)=\sqrt{2\pi}\,z^{2}(1-z^{2})e^{-z^{2}}$, and $\epsilon>0$
ensures convergence at small $y$. Now we take the logarithmic
derivatives of $p_{N}^{(2)}(bt)$ as in Eq.(\ref{d-log}) and
implement the limits $N/t\rightarrow\infty$ or
$t/N\rightarrow\infty$. The results are
\be \label{der}
-2\sqrt{2\pi}\tau^{\frac{\epsilon}{1-\epsilon}}\,\int_{0}^{\infty}\frac{dz}{z^{2}}\,{\cal
F}_{2}(z^{1-\epsilon}),\;\;-2\sqrt{2\pi}N^{\epsilon}\int_{0}^{\infty}\frac{dz}{z^{2}}\,{\cal
F}_{2}(z),
\en
respectively. Finally we take the limit $\epsilon\rightarrow 0$.
One can see that both expressions in Eq.(\ref{der}) coincide and
\be
\label{c1}
c_{1}=2\sqrt{2\pi}\int_{0}^{\infty}\frac{dz}{z^{2}}\,{\cal
F}_{2}(z)=\frac{\pi}{\sqrt{2}},
\en
provided that the operations of taking the limit $\epsilon\rightarrow 0$
and integrating over $z$ commute.

Thus the validity of the Chalker's ansatz, in the form given by Eq.(\ref{d-log}),
in the first order in $b\ll 1$ is based on the symmetry
of the integrand in Eq.(\ref{double-int}) with respect to $n$ and $y$
at $\epsilon=0$. However the necessary condition for this argument to
work is the {\it commutativity} of integrating up to infinity and taking
the limit $\epsilon\rightarrow 0$. Such a commutativity can be checked
straightforwardly for Eq.(\ref{der}).

Now we proceed with  $b^{2}$ contributions. Here the higher powers
of $\ln N$ or $\ln\tau$ may arise and their cancelation in $\ln
P_{N}(\tau)$ is of principal importance. The cancelation, if it
happens, excludes in Eq.(\ref{P(L)}) presence of further {\it
additive} terms such as $N^{-k\,d_{2}}$ or $\tau^{-k\,d_{2}}$ with
$k>1$. Note that such terms, albeit not changing the leading
large-$N,\tau$ behavior, could significantly alter the perturbative
expansion in $b\ln N$ and $b\ln\tau$. Their absence is a strong
argument in favor of a {\it pure power-law} behavior
Eqs.(\ref{P(L)},\ref{P(bt)}).

Using the representation for $p_{N}^{(3)}(bt)$ derived in Ref.\cite{YO07}
one can cast the combination ${\cal
P}_{N}^{(3)}(bt)=p_{N}^{(3)}(\tau)-\frac{1}{2}\,\left(p_{N}^{(2)}(\tau)\right)^{2}$
in the following form:
\begin{widetext}
\begin{eqnarray} \label{P-F} {\cal
P}_{\tau} \!\!\! & = & \!\!\! \tau^{\frac{2\epsilon}{1-\epsilon}}\left\{\left(
\frac{1-\epsilon}{2\epsilon}\right)\int \!\!\!\! \int_{-\infty}^{+\infty}dx
dy\,{\cal F}_{3}\left(x,y;\epsilon\right)- \frac{1}{2}\left(
\frac{1-\epsilon}{\epsilon}\right)^{2}\left[\int_{-\infty}^{+\infty}dx\,{\cal
F}_{2}\left( \frac{1}{|x|^{1-\epsilon}}\right)
\right]^{2}\right\} ,
 \\
 \label{P-F-N}
{\cal P}_{N} \!\!\! &=& \!\!\! N^{2\epsilon}\left\{ \! \int_{0}^{\infty}
  \!\!\!\! \frac{d\beta}{\beta^{1- 2\epsilon \kappa}}
\int \!\!\!\! \int_{\frac{-1}{\beta^{\kappa}}}^{\frac{1}{\beta^{\kappa}}}
 \!\!\! dx dy {\cal F}_{3}\left(x,y;\epsilon\right)\,\left(
1-|x-y|\beta^{\kappa}\right)
  -\frac{1}{2}\left[ \int_{0}^{\infty}
  \!\!\!\! \frac{d\beta}{\beta^{1-\epsilon\kappa }}
\int_{\frac{-1}{\beta^{\kappa}}}^{\frac{1}{\beta^{\kappa}}} \!\!\! dx {\cal F}_{2}\left(
\frac{1}{|x|^{1-\epsilon}}\right)\,\left(
1-|x|\beta^{\kappa}\right)\right]^{2} \! \right\}
\end{eqnarray}
where $ \, \kappa \equiv 1/(1 - \epsilon), \ {\cal P}_{N}=\lim_{\tau/N\rightarrow\infty}{\cal
P}_{N}^{(3)}(\tau)$ and ${\cal
P}_{\tau}=\lim_{N/\tau\rightarrow\infty}{\cal P}_{N}^{(3)}(\tau)$,
and
\begin{eqnarray} \label{F3}{\cal
F}_{3}\left(x,y;\epsilon\right)& = &-\frac{\sqrt{\pi} i}{8}\int_{{\cal
C}}d\alpha\,\sqrt{\alpha}\,e^{\alpha}\,\left[f_{1}(Z_{x})f_{1}(Z_{y})\,(f_{3}-f_{2})(Z_{x-y})+f_{2}(Z_{x})
f_{3}(Z_{y})\,(4f_{1}-f_{3})(Z_{x-y})\right],
   \\
\label{f-1-3}
f_{1}(Z)& = & Z/\sqrt{1+Z}, \ f_{2}(Z)=1/\sqrt{1+Z}, \ f_{3}(Z)=Z/(1+Z)^{\frac{3}{2}}, \
Z_{x}=1/(\alpha |x|^{2(1-\epsilon)}).
\end{eqnarray}
\end{widetext}
The contour ${\cal C}$ is the Hankel contour encompassing the
negative part of the real axis $ \, \Re (\alpha) < 0 $.

We present the cumbersome Eqs.(\ref{P-F})-(\ref{F3}) not only to
give a flavor of real complexity of the calculations but also
 to uncover the ultimate reason for the Chalker's ansatz to hold. As in
Eq.(\ref{der}), there is a certain similarity in Eqs.(\ref{P-F}) and
(\ref{P-F-N}) which can be traced back to $(n-m)^{-2}$ behavior
of the variance Eq.(\ref{var}). To exploit this similarity,
we use the approximate equality:
\begin{equation}\label{delta}
\int_{0}^{\infty} \!\!\!\! \frac{d\beta}{\beta^{1-\delta}} f(\beta) \simeq
\frac{f(0)}{\delta} - \int_{0}^{\infty} \!\!\!\! d\beta\,\ln\beta\,\frac{\partial
f}{\partial\beta}
-\frac{\delta}{2}\int_{0}^{\infty} \!\!\!\! d\beta\,\ln^{2}\beta\,\frac{\partial
f}{\partial\beta}.
\end{equation}
%
%
Applying this formula to  Eq.(\ref{P-F-N}), one can see that the first
term on the r.h.s. reproduces immediately Eq.(\ref{P-F}) up to the
change of the pre-factor $\tau^{\frac{2\epsilon}{1-\epsilon}}$ by
$N^{2\epsilon}$. Like in the dimensional regularization calculus
\cite{Iz-Zub}, after taking the logarithmic derivatives w.r.t. $\tau$
or $N$ and taking the subsequent limit $\epsilon\rightarrow
0$, the terms $\propto \epsilon^{-1}\,\tau^{\frac{2\epsilon}{1-\epsilon}}$
and $\propto \epsilon^{-1}\,N^{2\epsilon}$ tend to a constant
determining  the $b^{2}$ contribution to $d_{2}$.

The full calculation, however, is complicated by the presence of
$\epsilon^{-2}$ singularity in ${\cal F}_{3}$ leading to the appearance of
the additional contributions of the form  $\epsilon\, \epsilon^{-2}$ etc.
Thus one has to keep not only the first term on the r.h.s. of
Eq.(\ref{delta}) but the next two terms as well.
An accurate account of all such terms shows that Eq.(\ref{d-log}) is
valid to the $b^{2}$ order, and the fractal dimension $d_{2}$ is
equal to:
\begin{equation}
\label{d2} d_{2}=\frac{\pi b}{\sqrt{2}}+\frac{(\pi
b)^{2}}{4}\left[10-\frac{56}{3\sqrt{3}}-\ln4+\pi\,I\right]+O(b^{3}),
\end{equation}
where
\begin{equation}\nonumber
\label{I} I=\left( \frac{2}{\pi}\right)^{3} \!\!\! \int_{0}^{\frac{\pi}{2}} \!\!\!
\frac{d\varphi_{1}\,d\varphi_{2}\,d\varphi_{3}}{(\cos\varphi_{1}+
\cos\varphi_{2})\,
(\cos\varphi_{1}+\cos\varphi_{2}+\cos\varphi_{3})}.
\end{equation}
$ I \, $ can be evaluated numerically
\begin{equation}
\label{NI} I=0.79426047250532455983.
\end{equation}
However, we believe that this integral may have a geometrical
meaning which is still evading our comprehension and thus it can be
evaluated analytically. It remotely resembles the integrals arising
in the problem of resistance of the regular 3-dimensional resistor
lattice \cite{Levitov} which are known to have an intimate relation
with the number theory.

Note that the leading terms in $d_{2}(b)$ at small and large $b$
\cite{Mirlin-Rew} can be represented in the form of the duality
relation:
\begin{equation}
  \label{dual}
     d_{2}(B)+d_{2}(B^{-1})=1, \
     B \equiv ( \pi b ) 2^{\frac{1}{4}} .
\end{equation}
It appears that this relation is well fulfilled {\it at all} values of $B$
(see Fig.1).
\begin{figure}[t]
  \includegraphics[width=7cm, 
                               angle=0]{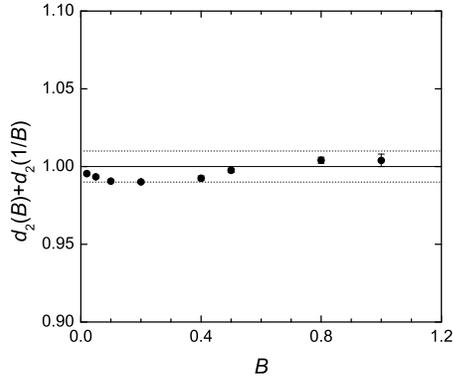}
  \caption{
   Duality relation verified by numerical diagonalization of
   RMT Eq.(\ref{var}). The deviation from Eq.(\ref{dual}) is less than
   $1\,\%$ for all values of $B=(\pi b)\,2^{\frac{1}{4}}$.
               }
\end{figure}
Our analytical result Eq.(\ref{d2}) together with the result
\cite{Wegner1987} which gives no $1/b^{2}$ terms in $d_{2}(b)$ for
large $b$, implies that the duality relation is not exact. However,
its extremely accurate {\it approximate validity} is only possible
because of the {\it anomalously small} value of the coefficient
$\approx 0.083$ in front of $(\pi b)^{2}$ in Eq.(\ref{d2}).

In conclusion, we have shown that the Chalker's dynamical scaling
and its drastic consequence for strong correlations of the sparse
multi-fractal wavefunctions is valid in the critical random matrix
ensemble in the limit of strong multi-fractality $d_{2}\ll 1$.
We checked its validity in the form of Eq.(\ref{d-log}) up to the
second order in the small parameter $b$ that controls the strength of
the multi-fractality. Specifically (i) we observed the
cancelation of the $\ln^{2}N$ and $\ln^{2}\tau$ terms in  $\ln
P_{N}(\tau)$ required by a pure power-law behavior;
(ii) we demonstrated that the coefficients in front of the $\ln N$
and $\ln\tau$ terms are the same up to the order $b^{2}$,  and (iii)
we found analytically the $b^{2}$ term in the fractal dimension
$d_{2}$. The validity of the Chalker's ansatz in the form Eq.(\ref{d-log})
is encoded in the possibility of symmetric representation of the two 
different limits which can be traced back to the $ (m-n)^{-2} $ 
dependence of the critical variance.

We acknowledge support from 1) the DFG through grant SFB
TR-12, and the Nanosystems Initiative Munich Cluster of Excellence
(OYe); 2) the Engineering and Physical Sciences Research Council,
grant No. EP/G055769/1 (AO); 3) FEDER and the Spanish DGI through
grant No. FIS2007-62238 (VEK, EC). OYe and AO acknowledge
hospitality of the Abdus Salam ICTP.

\end{document}